\documentclass[twocolumn,showpacs,preprintnumbers,amsmath,amssymb,prb]{revtex4}

\usepackage{graphicx}
\usepackage{dcolumn}
\usepackage{bm}

\begin{document}

\preprint{APS/123-QED}

\title{Cage-size control of guest vibration and thermal conductivity in Sr$_{8}$Ga$_{16}$Si$_{30-x}$Ge$_x$}

\author{K. Suekuni$^{1}$, M. A. Avila$^{1}$, K. Umeo$^{1,2}$, T. Takabatake$^{1,3}$}
\affiliation{%
$^{1}$Department of Quantum Matter, ADSM, $^{2}$Materials Science Center, N-BARD,\\
$^{3}$Institute for Advanced Materials Research, Hiroshima University, Higashi-Hiroshima 739-8530, Japan}

\date{\today}

\begin{abstract}
We present a systematic study of thermal conductivity, specific
heat, electrical resistivity, thermopower and x-ray diffraction
measurements performed on single-crystalline samples of the
pseudo-quaternary type-I clathrate system
Sr$_{8}$Ga$_{16}$Si$_{30-x}$Ge$_x$, in the full range of $0\leq
 x\leq 30$. All the samples show metallic behavior with
{\it n}-type majority carriers. However, the thermal conductivity and
specific heat strongly depend on $x$. Upon increasing $x$ from 0
to 30, the lattice parameter increases by 3\%, from 10.446 to
10.726 \AA, and the localized vibrational energies of the Sr guest
ions in the tetrakaidekahedron (dodecahedron) cages decrease from
59~(120)~K to 35~(90)~K. Furthermore, the lattice thermal
conductivity at low temperatures is largely suppressed. In fact, a
crystalline peak found at 15~K for $x = 0$ gradually decreases and
disappears for $x\geq 20$, evolving into the anomalous glass-like
behavior observed for $x = 30$. It is found that the increase of
the free space for the Sr guest motion directly correlates with a
continuous transition from on-center harmonic vibration to
off-center anharmonic vibration, with consequent increase in the
coupling strength between the guest's low-energy modes and the
cage's acoustic phonon modes.
\end{abstract}

\pacs{72.15.Jf, 72.20.Pa, 82.75.-z}

\maketitle

\section{\label{sec:int}INTRODUCTION}

Semiconducting clathrate compounds are attracting considerable
attention because of their potential for thermoelectric conversion
applications.\cite{nol01a} The efficiency of a thermoelectric
material at a given operation temperature $T$ can be quantified by
the dimensionless figure of merit $ZT =
S^{2}T/\rho(\kappa_{el}+\kappa_{L}$), where $S$, $\rho$,
$\kappa_{el}$, $\kappa_{L}$, are the thermopower, electrical
resistivity, electronic thermal conductivity and lattice thermal
conductivity of the material, respectively. Intermetallic
clathrates are compounds consisting of polyhedral cages (basically
formed by Si, Ge and Sn through diamond-like bonding) that are
normally filled with monovalent or divalent guest cations.\cite{eis86a,kov04a} 
Many of them follow the Zintl rule, where
the cage atoms are partially substituted by acceptor atoms for
charge compensation between guests and cages. In addition to the
large $S(T)$ and small $\kappa_{el}(T)$ often observed in Zintl
materials, the most pronounced feature of the clathrates is their
very low lattice thermal conductivity $\kappa_{L}$ (of order 1
W/m~K at room temperature).\cite{nol98a,cohn99a,nol00a,sal01a}
Some of these compounds even show glasslike temperature-dependant
thermal conductivity, although they crystallize in well-defined
structures. Therefore, clathrates are good candidates to fulfill
the phonon-glass electron-crystal (PGEC) concept,\cite{sla95a} 
which is a guideline to search for high-performance thermoelectric
materials with the very rare combination of simultaneously low
thermal conductivity and electrical resistivity. The determination
of which mechanisms are dominant in lowering $\kappa_{L}(T)$ in
clathrates presents motivation from the performance improvement
perspective, as well as from that of further understanding the
physics behind atoms vibrating in unconventional crystalline
lattices.

Among the several possible structures formed by these materials,
the type-I clathrate structure adopted by the
A$_{8}$Ga$_{16}$Ge$_{30}$ group (A = Ba, Sr and Eu) has shown
particularly favorable thermoelectric properties.
\cite{nol98a,cohn99a,nol00a,sal01a} The unit cell of this bcc
structure ($Pm\bar{3}n$, No. 223) consists of 46 cage atoms
arranged in two dodecahedra and six tetrakaidecahedra, which
incorporate two A atoms at the $2a$ site and six A atoms at the
$6d$ site, respectively.\cite{eis86a,kov04a} Initial
investigations of A$_{8}$Ga$_{16}$Ge$_{30}$ clearly indicated that
the $\kappa_{L}$ below room temperature was lowered in direct
relation to the decrease of the guest atom's ionic radius
(Ba$^{+2}$ largest, Eu$^{+2}$ smallest), whereas the lattice
parameter of the three compounds remains relatively unchanged.
\cite{sal01a} This suggested that the general lowering in thermal
conductivity is not so much related to the mass of the guest ions,
but rather realized by the fact that they are loosely bound to an
oversized cage, giving rise to an anomalous vibration showing
low-frequency, non-dispersive localized modes (rattling) which
couple to the heat-carrying acoustic phonon modes of the rigid
cage structure and scatter them efficiently.
\cite{nol00a,nol00b,dong00a,dong01a,taka06b}

\begin{table*}[bt]
\caption{Starting (flux) composition, crystal composition, lattice
parameter $a$, electrical resistivity $\rho$, thermopower $S$ and
carrier concentration $n$ at room temperature of
Sr$_{8}$Ga$_{16}$Si$_{30-x}$Ge$_x$}.
\begin{ruledtabular}
\begin{tabular}{ccccccccccccc}
\\
$\ $ & starting composition & Sr & : & Ga & : & Si & : & Ge & $a$ (\AA) & $\rho$$_{280K} $(m$\Omega$~cm) & $S$$_{280K}$ ($\mu$V/K) & $n$ (10$^{20}$/cm$^{3}$)\\
\\
\hline
\\
$x$=0 & 8 : 38 : 30 : 0 & 8 & : & 13.6 & : & 32.4 & : & 0 & 10.446 & 0.26 & -13 & 46 \\
5 & 8 : 38 : 24 : 6 & 8 & : & 13.7 & : & 27.0 & : & 5.3 & 10.483 & 0.28 & -26 & 40 \\
20 & 8 : 38 : 15 : 15 & 8 & : & 15.7 & : & 10.0 & : & 20.3 & 10.638 & 0.68 & -60 & 16 \\
26 & 8 : 38 : 6 : 24 & 8 & : & 15.9 & : & 4.0 & : & 26.1 & 10.703 & 1.23 & -133 & 10 \\
30 & 8 : 38 : 0 : 30 & 8 & : & 15.9 & : & 0 & : & 30.1 & 10.726 & 1.85 & -200 & 4 \\
\\
\end{tabular}
\end{ruledtabular}
\end{table*}

In addition, $\kappa$($T$) for Ba$_{8}$Ga$_{16}$Ge$_{30}$ exhibited a large peak at 15~K, being
characteristic of a crystal lattice.\cite{sal01a} By contrast,
$\kappa$($T$) for Sr$_{8}$Ga$_{16}$Ge$_{30}$ and
Eu$_{8}$Ga$_{16}$Ge$_{30}$ showed all of the characteristics of a
structural glass.\cite{nol98a,nol00a,sal01a} Neutron diffraction
measurements\cite{sal01a} revealed that the Ba atom in the large
cage is located essentially at the center of the cage ($6d$ site),
whereas a substantial probability exists for the Sr atom to move
off the site center about 0.3~\AA, to one of four
crystallographically equivalent positions ($24j$ or $24k$ sites),
and Eu atoms move away even more, 0.4~\AA~ from the $6d$ site,
suggesting that off-center rattling may be necessary to produce
glass-like thermal conductivity
.\cite{bri04a,baum05a,mad05a,herm05a} At lower temperature,
nuclear tunneling among the four sites
\cite{zer04a,gou05a,herm06a} may also play the role. The on-center
vibrational freedom of the Ba ions can be adequately described
assuming independent harmonic oscillators (Einstein model),
\cite{sal01a,bri04a,ben04a,umeo05a,avi06c} but the Sr and Eu
vibration cannot be satisfactorily modeled this way, indicating
that anharmonic vibration contributions gain significance in these
cases.\cite{bri04a} For Sr$_{8}$Ga$_{16}$Ge$_{30}$, the anomalous
specific heat contribution of the Sr ions was successfully
reproduced by using a soft-potential model.\cite{umeo05a} The
appearance of a plateau in $\kappa(T)$ and a broadened maximum in
$C/T^{3}$ were interpreted as resultant from the low-energy
excitations associated with anharmonic, quasi-localized vibrations
of the Sr ions in the cage.

As an added degree of complexity, later investigations on several
Ba-based clathrates have shown a strong dependence of the thermal
conductivity on the majority charge carrier type,
\cite{ben04a,avi06a} such that compounds with {\it n}-type carriers
show low-temperature crystalline peaks in $\kappa(T)$ while their
{\it p}-type counterparts show significantly lower and glasslike
$\kappa(T)$ in the same temperature interval.
\cite{ben04a,ben06a,avi06a,avi06b,avi06c} These results appear
inconsistent with the idea of the guest ion vibration having the
relevant role in producing the glass-like behavior. An alternate
model\cite{ben04a,ben05a,pach05a,ben06a} based on
phonon-scattering by charge carriers\cite{ziman60a} was proposed
instead, and the question of which factors are dominant at low
temperatures remains a currently open debate.\cite{avi06c}

Given this scenario, it seemed instructive to pursue an
investigation of the guest vibrational behaviors and their effect
on thermal conductivity from the opposite approach, by
maintaining the guest atom and carrier type fixed, while varying
the cage environment in a controlled and systematic manner. The
compound Sr$_{8}$Ga$_{16-y}$Si$_{30+y}$ ($0\leq y\leq 5$) has only
been characterized structurally so far
\cite{eis86a,imai02a,cab02a} and was reported to adopt the same
type-I clathrate structure as A$_{8}$Ga$_{16}$Ge$_{30}$. The
lattice parameter is somewhat dependent on $y$ due to the size
difference between Ga and Si, but remains between 10.460~\AA ~
($y=0$) and 10.408~\AA ~($y=5$), therefore about 3\% smaller than
that of Sr$_{8}$Ga$_{16}$Ge$_{30}$.\cite{eis86a} Furthermore,
partial solid solutions of Si-Ge in a clathrate structure have
already been realized in polycrystalline
Ba$_{8}$Ga$_{16}$Si$_{30-x}$Ge$_x$.\cite{nak02a, mart06a} Based
on all the aforementioned, the pseudo-quaternary system
Sr$_{8}$Ga$_{16}$Si$_{30-x}$Ge$_x$ appears to offer a suitable
opportunity for investigating the relationships between cage size,
guest vibration and thermal conductivity. Indeed, we will show
that homogeneous single-crystalline samples within the full range
$0\leq x\leq 30$ can be achieved and, upon increase of $x$ and
expansion of the cage size, we find a continuous evolution from
on-center harmonic vibration and crystal-like $\kappa_L(T)$, to
off-center anharmonic vibration and glass-like $\kappa_L(T)$.

\section{\label{sec:exp}EXPERIMENTAL}

Polyhedral single crystals of 1-5 mm in diam. were grown from a
self-flux method using excess Ga. High purity elements were mixed
in an atomic ratio of Sr:Ga:Si:Ge = 8:38:(30-X):X (X = 0, 6, 15,
24, 30) in an argon filled glovebox. The mixture was sealed in an
evacuated and carbonized quartz tube, soaked at 1180~$^{\circ}$C
for 2-3~h, cooled over 10~h to 1000$^{^{\circ}}$C and then slowly
cooled over 100~h to 800-700$^{^{\circ}}$C. At this point the
ampoules were quickly removed from the furnace and the remaining
molten Ga flux was separated by centrifuging. The composition of
the crystals was examined by electron probe microanalysis (EPMA)
with a wave length dispersive JEOL JXA-8200 system. The results
are shown in Table I. The compositions of Si and Ge in the
crystals are somewhat different from the starting compositions,
and Ga is deficient in the Si rich crystals. Hereafter, the Ge
content in the crystal $x$ is used to denote the
Sr$_{8}$Ga$_{16}$Si$_{30-x}$Ge$_x$ samples.

\begin{figure}[tb]
\includegraphics[width=8cm,clip]{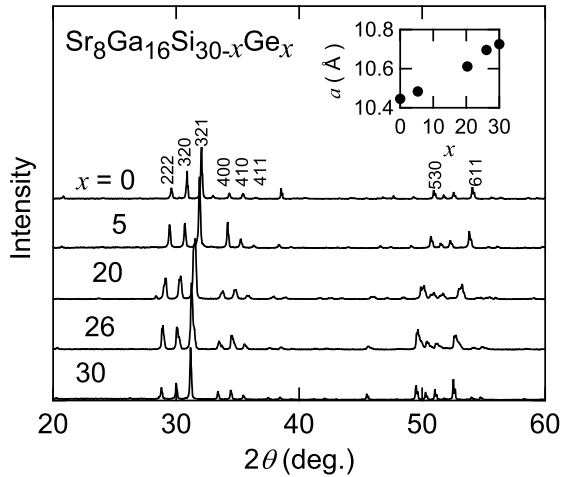}
\caption{\label{fig:xrd}Powder x-ray diffraction spectra (Cu
$K\alpha$ radiation) for the Sr$_{8}$Ga$_{16}$Si$_{30-x}$Ge$_x$
specimens, and refined lattice parameters (inset).}
\end{figure}

The crystal structures were refined with powder x-ray diffraction
(XRD). The spectra of all
samples shown in Fig.~\ref{fig:xrd} were indexed on the basis of 
the type-I clathrate structure. Since both the atomic
size and bond length of Ge are larger than for Si, the lattice
parameter linearly increases as $x$ increases from 0 to 30 in the
series of samples, such that there is a 3\% difference between the
end compounds (see inset). For $x = 20$ and 26, higher angle peaks
like [530] and [611] are visibly broadened, possibly due to the
disorder among Si, Ga and Ge atoms. Electrical resistivity,
thermopower and Hall coefficient were measured in home-made
systems by a standard DC four-probe method, differential method
and DC technique, respectively in the temperature range from 4 to
300~K. Thermal conductivity experiments were performed using a
steady-state method in a home-made cryostat. The data is reliable
up to about 150~K, above which the effect of thermal losses by
radiation and wire conduction require corrections. The specific
heat from 0.3 to 300~K was measured using a Quantum Design PPMS
with a thermal-relaxation method.

\section{\label{sec:res}RESULTS AND DISCUSSION}

\begin{figure}[htb]
\includegraphics[width=8cm,clip]{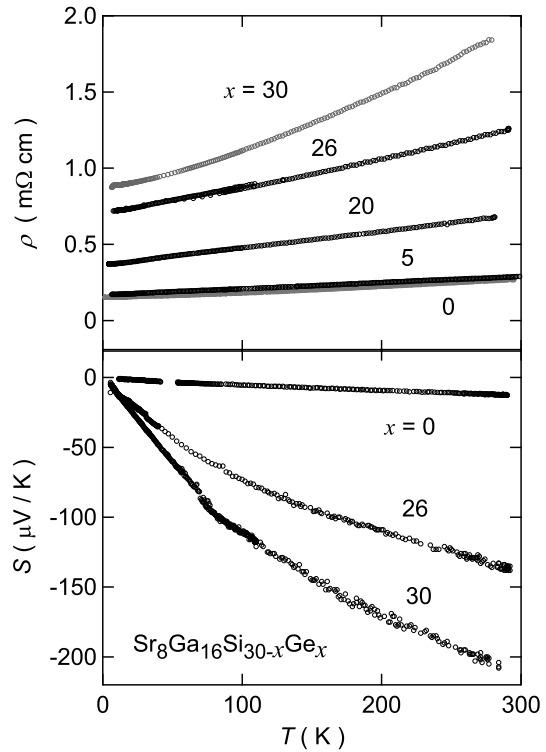}
\caption{\label{fig:transp}Temperature dependence of electrical
resistivity $\rho$ and thermopower $S$ for
Sr$_{8}$Ga$_{16}$Si$_{30-x}$Ge$_x$.}
\end{figure}

The values of electrical resistivity ($\rho$), thermopower ($S$),
and carrier concentration ($n$) at room temperature are listed in
Table I. The latter was estimated from the Hall coefficient
assuming one type of carriers. The decreasing trend in $n$ with
increasing $x$ is consistent with the systematic increase of
$\rho$ ($T=280$~K) from 0.26 to 1.85~m$\Omega$~cm. The temperature
dependences of $\rho$ and $S$ are shown in Fig.~\ref{fig:transp}.
The monotonic decrease of $\rho$($T$) upon cooling is characteristic of heavily
doped semiconductors or low carrier density metals. For all
samples, $S$ is negative and the absolute value increases
monotonically with the increase of $x$. These variations of the
transport properties can be attributed to a systematic decrease in
$n$, from $46\times~10^{20}$/cm$^{3}$ to
$3.6\times~10^{20}$/cm$^{3}$. This arises from the fact that
Si-rich samples tend to be Ga deficient (see Table I and
refs.~\onlinecite{eis86a,imai02a,cab02a}) while Ge-rich samples
follow the Zintl rule more strictly, and are therefore less
electron-doped than Si-rich samples.

\begin{figure}
\includegraphics[width=8cm,clip]{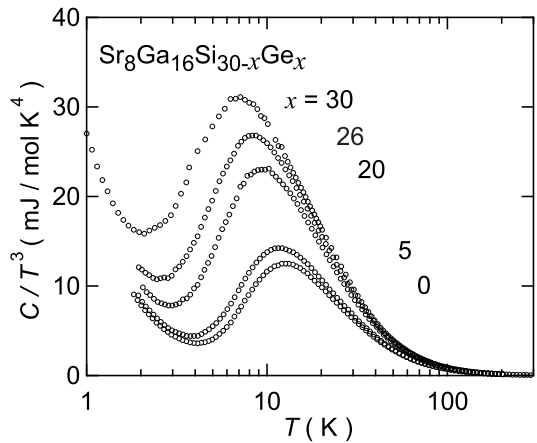}
\caption{\label{fig:heat}Temperature dependence of specific heat
$C$ for Sr$_{8}$Ga$_{16}$Si$_{30-x}$Ge$_x$, presented as
$C/T^3$~vs~$T$.}
\end{figure}

The specific heats are plotted as $C/T^{3}$~vs.~$T$ in
Fig.~\ref{fig:heat}. In this plotting style, the contribution of
rattling ions appears as a broad peak over an electronic and Debye
``background'' from the stiff cage. Upon cooling, {\it C/T}$^{3}$
for all samples initially rises into the broad peak, followed by a
local minimum, and finally rises again as 1/$T$$^{2}$ due to the
contribution from conduction electron and/or tunneling of guest
ions. With increasing $x$, the peak height rises systematically
and its temperature shifts from 10.5~K to 7~K. This is not related
to $n$, whose contribution becomes vanishingly small at
temperatures above 4~K. Rather, it is already a qualitative
demonstration of how the rattling of the Sr ions increases in
direct relation to the cage size. Further information about the Sr
vibration characteristics can be obtained through a more careful
analysis of the data as follows, using the same methodology we
previously developed to analyze the data of {\it C} of
Ba$_{8}$Ga$_{16}$Ge$_{30}$.\cite{avi06c}

As a first approximation, the Sr atoms can be considered
independent Einstein oscillators, and the framework composed of
(Ga,Si,Ge)$_{46}$ cages a stiff Debye solid. Following this
approach, $C$ of Sr$_{8}$Ga$_{16}$Si$_{30-x}$Ge$_x$ is treated as
a sum of three terms: an electronic contribution $C_{el}$, a Debye
contribution $C_{D}$, and an Einstein contribution $C_{E}$. As we
described in detail in ref.~\onlinecite{avi06c}, first the
Sommerfeld coefficient $\gamma$ and the Debye temperature
$\theta_{D}$ should be evaluated independently and fixed, together
with the pre-defined dimensionalities and numbers of Einstein
oscillators, so that only two fitting parameters are left, which
characterize the Sr guests vibrational energies. It should be
recalled that the six Sr$_{(6d)}$ ions show a strongly anisotropic
vibration with greater amplitude within the plane parallel to the
larger cage's two hexagons.\cite{nol00a} Because the
dimensionality plays a role in the Einstein model, at least two
vibrational modes should be required to describe the $6d$ site
alone: in-plane $\theta_{E(6d)}^{\parallel}$ and out-of-plane
$\theta_{E(6d)}^{\perp}$, respectively.\cite{dong00a} In
addition, a third vibrational mode $\theta_{E(2a)}$ is required to
account for the smaller, but still Einstein-type rattling of the
two Sr$_{(2a)}$ which can be assumed isotropic.\cite{sal01a} In
our model, the dimensionalities and numbers of oscillators are
predefined: $N_{E(2a)} = 3\times2$, $N_{E(6d)}^{\parallel} =
2\times6$, $N_{E(6d)}^{\perp} = 1\times6$. We impose the
additional constraints
$\theta_{E(6d)}^{\parallel}<\theta_{E(6d)}^{\perp},\theta_{E(2a)}$,
$\theta_{E(6d)}^{\parallel} = \theta_{EL}$ and
$\theta_{E(6d)}^{\perp} = \theta_{E(2a)} = \theta_{EH}$, so the
fitting parameters are only the lower and higher Einstein
temperatures, $\theta_{EL}$ and $\theta_{EH}$ respectively.

From $C/T$~vs.~$T^{2}$ plots (not shown) the obtained values of
$\gamma$ are between 11 and 24~mJ/mol~K$^{2}$, and the $\theta_{D}$ 
decreases from 370~K to 200~K on going from $x = 0$ to
30. However, as we discussed previously,\cite{umeo05a} when
Sr$_{(6d)}$ anharmonic vibration becomes relevant it interferes
with this evaluation even at the lowest temperatures. In fact,
this last value of 200~K for $\theta_{D}$ of Sr$_{8}$Ga$_{16}$Ge$_{30}$
is an artifact, much smaller than
312~K estimated from atomic displacement parameters.\cite{ben05a}
Thus, we use the value of 288~K obtained for
Ba$_{8}$Ga$_{16}$Ge$_{30}$ as a better representation of the Ga-Ge
cages' Debye temperature for Sr$_{8}$Ga$_{16}$Ge$_{30}$. For
$x = 5$, 20, 26, $\theta_{D}$ is estimated by linear interpolation
between the values for $x = 0$ and $x = 30$.

\begin{figure}
\includegraphics[width=8cm,clip]{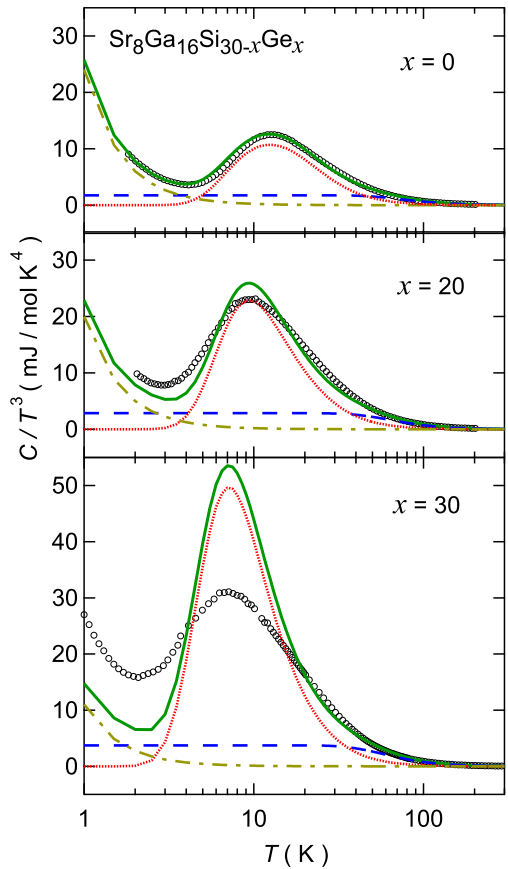}
\caption{\label{fig:einstein}(Color online) Einstein model fits
(green solid lines) of the $C/T^3$ data (symbols). The dotted,
dashed and dashed-dotted lines are the calculated $C_E$, $C_D$,
and $C_{el}$, respectively (see text).}
\end{figure}

Fig.~\ref{fig:einstein} shows the fits to the data of $C/T^{3}$
for $x = 0$, 20 and 30. For $x = 0$, an excellent fit is obtained
with $\theta_{EL}$ = 59~K and $\theta_{EH}$ = 120~K, indicating
that the assumption of independent, harmonic oscillators contained
within the Einstein model applies very well for the Sr vibration
in Sr$_{8}$Ga$_{16}$Si$_{30}$. By $x = 20$ it is clear that the
fit with the Einstein model is no longer a good representation for
the data, and at $x = 30$ the model has become completely
inadequate to reproduce the behavior. This is a clear indication
of the effects caused by increasingly anharmonic vibration of the
Sr$_{(6d)}$ guest ions, as their potential well broadens in direct
proportion to the free space available for movement
.\cite{dong00a} In such a case, a soft potential model gives a good
fit to $C$($T$).\cite{umeo05a} To our
knowledge, however, there are no theoretical models available that
can describe the $C$($T$) in this continuous transition from
harmonic to anharmonic vibration observed in our sample series.
For $x = 30$, the value of $\theta_{EL} = 35$~K determined mainly
by the peak position is in good agreement with the result of
 38~K observed by Raman scattering.\cite{taka06b} The
Einstein temperatures $\theta_{EL}$ and $\theta_{EH}$ for all
samples are listed in Table II.

\begin{figure}
\includegraphics[width=8cm,clip]{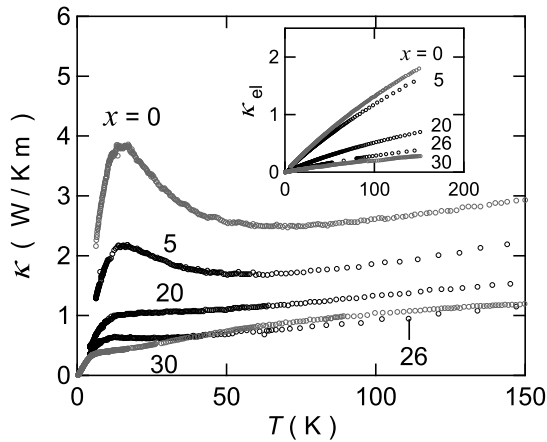}
\caption{\label{fig:kappa}Temperature dependence of total thermal
conductivity $\kappa$. The inset shows the estimated electronic
part $\kappa_{el}$.}
\end{figure}

Let us now see how all of these systematic changes affect the heat
transport, which is our main purpose. The total thermal
conductivity $\kappa$ of Sr$_{8}$Ga$_{16}$Si$_{30-x}$Ge$_x$ is
plotted as a function of temperature up to 150~K in
Fig.~\ref{fig:kappa}. We can see two types of systematic
evolutions in the data with increasing $x$: (i) $\kappa$ at higher
temperature decreases by a factor of 3, and (ii) a low temperature
crystalline peak for $x = 0$ is gradually but strongly suppressed,
resulting in glasslike behavior for $x = 30$. The first effect can
be directly attributed to the electronic contribution
$\kappa_{el}(T)$, which is estimated from the electrical
resistivity $\rho(T)$ using the Wiedemann-Franz law,
$\kappa_{el}(T)=(\pi^{2}k_{B}^{2}/3e^{2})T/\rho(T)$. As shown in
the inset of Fig.~\ref{fig:kappa}, $\kappa_{el}(T)$ have a simple,
slightly sublinear behavior with slopes directly related to the
carrier concentration, and therefore no relevant effect on the low
temperature peak.

\begin{figure}
\includegraphics[width=8cm,clip]{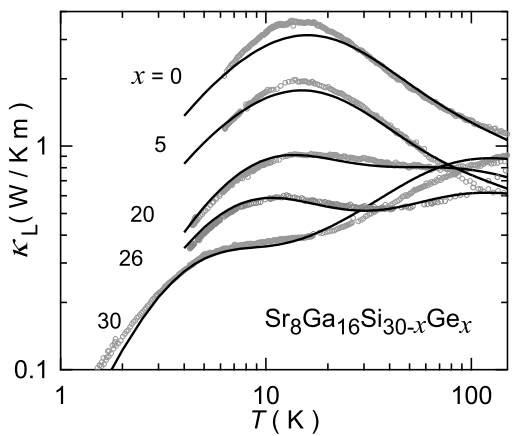}
\caption{\label{fig:kappaL}Temperature dependence of lattice
thermal conductivity $\kappa_{L}$. The solid lines are fits using
the TRR model described in the text.}
\end{figure}

By subtracting $\kappa_{el}(T)$ from $\kappa(T)$, we can 
estimate the lattice contributions $\kappa_{L}(T)$, which are
shown in Fig.~\ref{fig:kappaL}. The values are now much closer at
temperatures above 100~K, and an interesting feature reveals
itself in this range: the heat conduction level of the
pseudo-quaternary (intermediate) samples is lowered with respect
to the ternary (end) samples. This is most likely the effect of
extra phonon scattering on Ga/Si/Ge site disorder on the cage.

At lower temperatures, the plot shows a clearer picture of the
peak suppression. We will analyze the data using the same approach
described in detail in ref.~\onlinecite{avi06c}. For lack of a
single model capable of describing all sample behaviors in $C$($T$), we
use the results of the Einstein model fittings despite their poor
quality in Ge-rich samples. This is justifiable in our comparative
analysis because $\kappa_{L}(T)$, given in the semiclassical
theory by

\begin{eqnarray}
\kappa_L=\frac{1}{3}\int_0^{\omega_D} d\omega
\left[C_L(\omega,T)\nu l \right],
\end{eqnarray}

is not limited by the lattice specific
heat $C_{L}(\omega,T)$, nor the average sound velocity $v$, but
rather by the very low phonon mean free path $l$, which is
averaged over all major contributing scattering mechanisms. In the
TRR model,\cite{cohn99a} $l$ is written as

\begin{eqnarray}
l=(l^{-1}_{TS}+l^{-1}_{res}+l^{-1}_{Ray})^{-1} + l_{min}
\end{eqnarray}

which includes three mechanisms: tunneling between localized guest
sites, resonance scattering from guest ion rattling and Rayleigh
scattering from impurities, imperfections and mass difference.

The best fits to the data are shown as solid lines in
Fig.~\ref{fig:kappaL} and the parameter values are summarized in
Table II. The most relevant results are the increase by one order of magnitude in both the
resonant scattering level $C_{i}$ and the TS scattering level as
$x$ increases. The latter can be expressed by the ratio $A/B =
\tilde{n}(\hbar \nu)^2/k_B$, which in glasses is essentially a
measure of subset density of tunneling states $\tilde{n}$ that are
able to strongly couple to the phonons and effectively scatter
them.\cite{grae86a} For $x = 5$, 20, 26, Si disorder among Ga and
Ge makes Rayleigh scattering level larger than that for $x = 0$
and 30.

\begin{table}
\caption{Parameters used for the solid line curves in
Fig.~\ref{fig:einstein} and Fig.~\ref{fig:kappaL}.}
\begin{ruledtabular}
\begin{tabular}{ccccccc}
\\
Parameter & Unit & $x$=0 & 5 & 20 & 26 & 30\\
\\
\hline
\\
$\theta_{EL}$ & K & 59 & 56 & 45.5 & 40.5 & 35\\
$\theta_{EH}$ & K & 120 & 115 & 100 & 94 & 90\\
$C_1$ & $10^{30}$/(m s$^2$ K$^2$) & 0.3 & 0.5 & 1.3 & 2.1 & 4.2\\
$C_2$ & $10^{30}$/(m s$^2$ K$^2$) & 0.1 & 0.16 & 0.4 & 0.7 & 1.4\\
$A/B$ & 10$^5$/(m K) & 0.65 & 1.1 & 3.5 & 4 & 4.5\\
$D$ & 1/(m K$^4$) & 1.5 & 3 & 2 & 3 & 1.5\\
$\theta_D$ & K & 370 & 355 & 315 & 300 & 288\\
\\
\end{tabular}
\end{ruledtabular}
\end{table}

\begin{figure}
\includegraphics[width=8cm,clip]{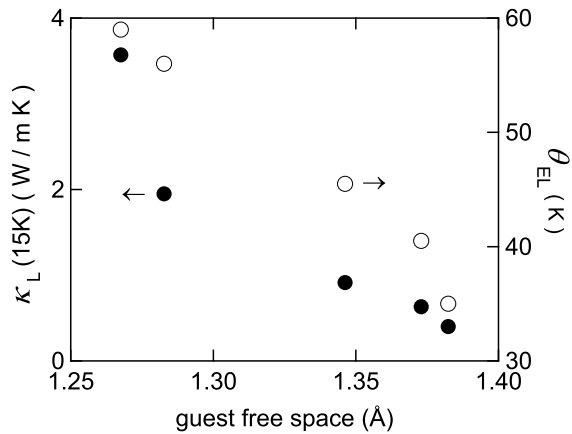}
\caption{\label{fig:freespace}Dependence of the lower Einstein
temperature $\theta_{EL}$ and lattice thermal conductivity at 15~K
on the guest free space.}
\end{figure}

Thus, the combined results of lattice $C_L$($T$) and $\kappa_L$($T$)
can be clearly attributed to a systematic evolution of the
Sr$_{(6d)}$ guests rattling level, with a decrease in
characteristic energy and increase in anharmonicity, both arising
from the fact that the lattice expansion increases the free space
for guest excursion and deforms the restoration potential. We may
conclude that this is the main cause of the peak suppression in
$\kappa_L$, leading to glasslike behavior.
Fig.~\ref{fig:freespace} summarizes these results by plotting
$\kappa_L$ at 15~K and $\theta_{EL}$ in terms of the guest free
space, evaluated semi-quantitatively by subtracting the Sr ionic
radius (1.35~\AA) and Ga covalent radius (1.26~\AA) from the
tetrakaidecahedron's radius in each sample.

\section{\label{sec:con}CONCLUSION}

By growing homogeneous single crystals of
Sr$_{8}$Ga$_{16}$Si$_{30-x}$Ge$_{x}$ in the full range of 0 $\leq$
$x$ $\leq 30$, we were able to gain systematic control of a type-I
clathrate structure's cage size without changing the guest ion or
the charge carrier type. The lattice parameter expands by up to
3\% with increasing Ge content, and as a consequence the free
space for guest excursion increases in the cage. The
characteristic energy of the localized Sr$_{(6d)}$ vibrations
decreases from 59~K to 35~K, and the Sr$_{(6d)}$ behavior clearly
evolves from vibrating in an on-center harmonic potential in
Sr$_{8}$Ga$_{16}$Si$_{30}$ to a broadened potential in
Sr$_{8}$Ga$_{16}$Ge$_{30}$, which allows off-center and anharmonic
vibration. This leads to an increase in the effective density of
tunneling states and a strong enhancement of the coupling between
the Sr$_{(6d)}$ vibration and the cage acoustic phonons,
shortening the latter's mean free path. As a result, the low
temperature (1-20~K) lattice thermal conductivity is suppressed in
such a way that the crystalline-like peak found for
Sr$_{8}$Ga$_{16}$Si$_{30}$ evolves into the well-known glasslike
behavior of Sr$_{8}$Ga$_{16}$Ge$_{30}$.

Our results leave little doubt that the described mechanism is the
dominant one in producing the anomalous thermal conductivity
behaviors observed in these clathrates at low-intermediate
temperatures. However, other factors certainly contribute, with
variable relevance depending on the particular system or
temperature range under study.

\begin{acknowledgments}
We thank Y. Shibata for the electron probe microanalysis performed at
N-BARD, Hiroshima University. This work was financially supported by
the Grants in Aid for Scientific Research (A) (Grant No. 18204032)
and the priority area "Skutterudite" (Grant No. 15072205) from MEXT, Japan.
\end{acknowledgments}


\end{document}